# Ultra-Low-Loss Silicon Nitride Photonics Based on Deposited Films Compatible with Foundries

*Xingchen Ji,[1,3,*] Yoshitomo Okawachi,[2] Andres Gil-Molina,[1] Mateus Corato-Zanarella,[1] Samantha Roberts,[1] Alexander L. Gaeta,[2] and Michal Lipson[1,*]*

[1]Department of Electrical Engineering, Columbia University, New York, NY, 10027, USA

[2]Department of Applied Physics and Applied Mathematics, Columbia University, New York, NY, 10027, USA

[3]Currently at John Hopcroft Center for Computer Science, School of Electronic Information and Electrical Engineering, Shanghai Jiao Tong University, Shanghai 200240, China

[*]Corresponding Author: E-mail: xingchenji@sjtu.edu.cn and ml3745@columbia.edu

**Abstract:**

The fabrication processes of silicon nitride ($Si_3N_4$) photonic devices used in foundries require low temperature deposition, which typically leads to high propagation losses. Here, we show that propagation loss as low as 0.42 dB/cm can be achieved using foundry compatible processes by solely reducing waveguide surface roughness. By post-processing the fabricated devices using rapid thermal anneal (RTA) and furnace anneal, we achieve propagation losses down to 0.28 dB/cm and 0.06 dB/cm, respectively. These low losses are comparable to the conventional devices using high temperature, high-stress LPCVD films. We also tune the dispersion of the devices, and proved that these devices can be used for linear and nonlinear applications. Low threshold parametric oscillation, broadband frequency combs and narrow-linewidth laser are demonstrated. Our work demonstrates the feasibility of scalable photonic systems based on foundries.

## 1. Introduction

To date, ultra-low-loss silicon nitride ($Si_3N_4$) waveguides and resonators have been demonstrated almost exclusively using films deposited at high temperature, while foundries mostly rely on $Si_3N_4$ films deposited at low temperature. The high temperature deposition uses low-pressure chemical vapor deposition (LPCVD), while low temperature deposition uses plasma-enhanced chemical vapor deposition (PECVD). PECVD $Si_3N_4$ is the most commonly used thin film in foundries as an insulator or a chemical barrier layer, however, the high propagation losses in these films limit their applications in photonics. LPCVD $Si_3N_4$ is not used in foundries due to the high temperature required and high film stress. Therefore, reducing losses in PECVD $Si_3N_4$ photonic devices is critical for integrating photonics devices with electronics, which could be used to realize high performance, scalable systems and realize system-level innovation[1].

Previously, there have been efforts to reduce losses in PECVD $Si_3N_4$ films by chemically changing the film composition[2–5]. By lowering the ammonium concentration during the deposition, losses down to 1.5 dB/cm have been shown[2]. However, these losses remain too high for most photonic applications. Researchers have also substituted conventional precursors with deuterated ones to reduce the losses of the film, losses down to 0.3 dB/cm have been shown[6]. However, these methods require special precursors and deposition tools, which are not commonly available in foundries.

## 2. Film deposition and waveguide fabrication

Here we show that low-loss can be achieved in a standard PECVD process by physically reducing waveguide surface roughness. The fabrication process is schematically shown in **Figure 1**. We deposit $Si_3N_4$ using PECVD at 350 °C in a single step onto a thermally oxidized 4-inch silicon wafer. The gases used for deposition are a mixture of silane (SiH$_4$: 20 sccm) diluted by nitrogen (N$_2$: 1425 sccm) and pure ammonia (NH$_3$: 30 sccm), with a process pressure of 1900 mTorr. The plasma frequencies alternate between a high frequency (13.56 MHz) with a power of 200 W and a low frequency (100 kHz) with a power of 160 W. The time duration for the two frequencies is 8 seconds and 12 seconds, respectively. The above parameters ensure that the deposition of Si$_3$N$_4$ film has very low film stress and high uniformity. The measured stress for the Si$_3$N$_4$ film on a test wafer is 93.4 MPa and tensile, which is more than an order of magnitude lower than LPCVD Si$_3$N$_4$ films deposited at high temperature. The low stress allows us to deposit thicker films without any cracking.

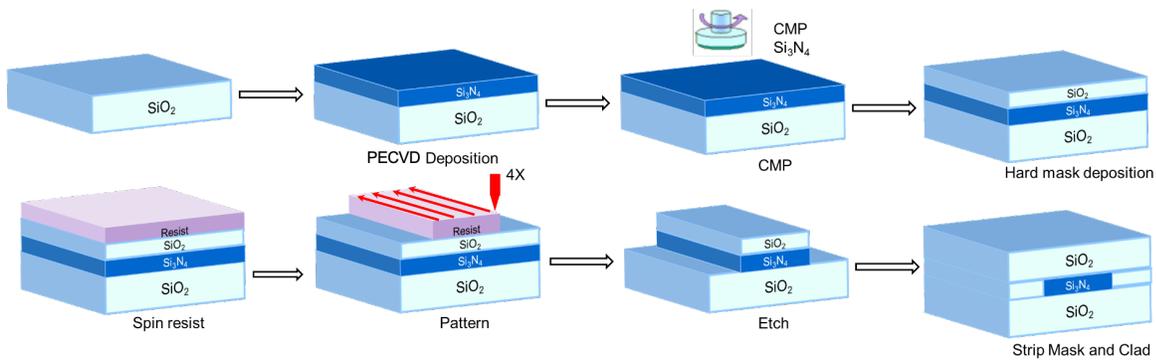

**Figure 1**. Schematic of our low-temperature PECVD Si$_3$N$_4$ fabrication processes. The process steps here are fully compatible with CMOS electronics.

We design high confinement waveguides based on the deposited PECVD films allows for strong dispersion engineering. One can see in **Figure 2**, the strong mode overlaps with the top surface that can exhibit a roughness of several nanometers for PECVD films[7,8].

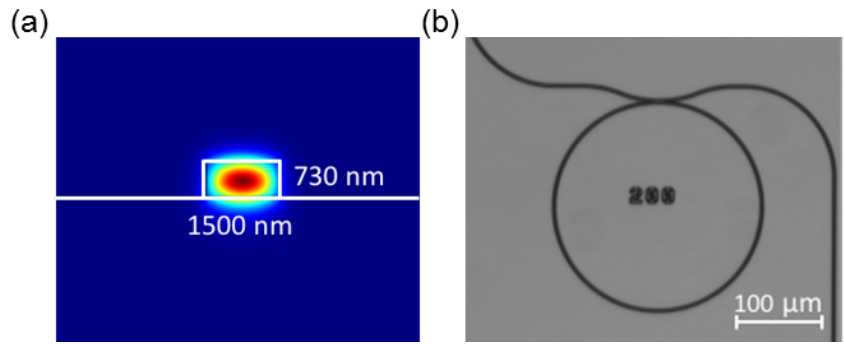

**Figure 2**. Mode simulation and microscope images of fabricated devices. (a) Mode simulation of 730 nm tall and 1500 nm wide waveguide showing that the mode is highly confined in the geometry we have chosen. (b) Top view optical microscope image of a 115 μm radius ring resonator.

To reduce scattering from the top surface of PECVD $Si_3N_4$, we use chemical mechanical planarization (CMP) to smooth the surface, as roughness traditionally leads to a high loss. We show the atomic-force microscopy (AFM) scans before and after the polishing step in **Figure 3**. The root-mean-squared (RMS) roughness is decreased from 1.36 nm before polishing to 0.20 nm after polishing. In order to reduce the roughness from the sidewalls and protect the polished top surface, we use a $SiO_2$ hard mask deposited using PECVD after CMP and use a dry etching process with a much higher oxygen flow. This etching process has been proved to substantially reduce the polymerization process during etching and decreases the roughness[9]. We pattern our devices with electron beam lithography using ma-N 2403 resist and use multipass writing algorithms to further reduce sidewall roughness caused by the lithography itself[9,10]. Finally, we clad the devices with 2 μm of $SiO_2$ deposited using PECVD for waveguide protection. The fabricated devices consist of resonators with a radius of 115 μm, a height of 730 nm and a width of 1500 nm, which are coupled to a waveguide of the same width and height. These dimensions ensure high confinement.

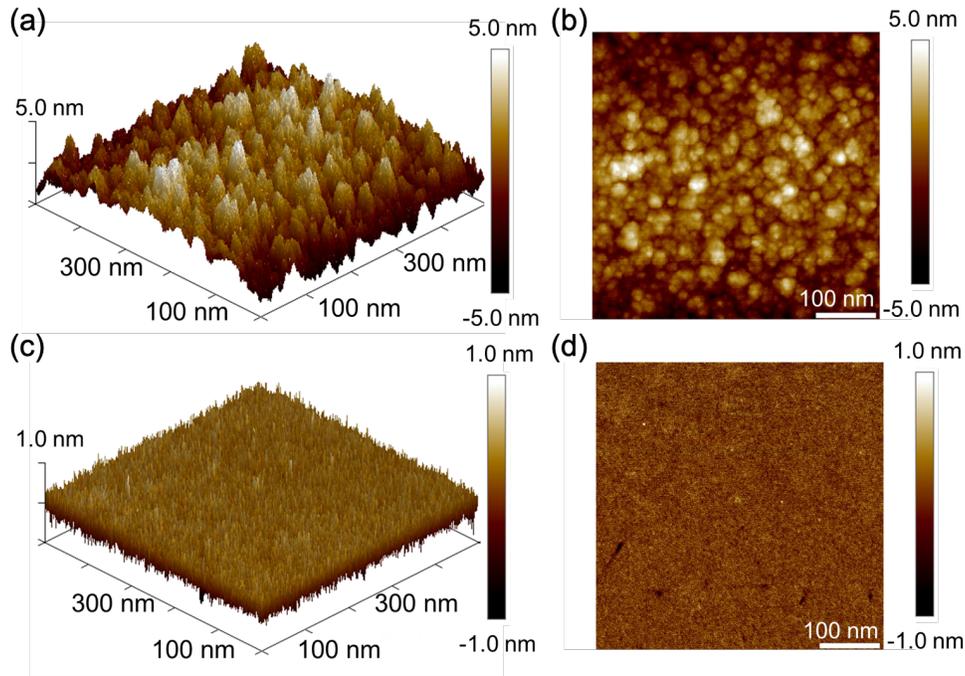

**Figure 3**. AFM measurement of the top surface of PECVD Si₃N₄. (a) 3D AFM scan of the top surface before CMP with RMS roughness of 1.36 nm and a correlation length of 27.6 nm. (b) 2D image of Si₃N₄ top surface before CMP and scaled to -5.0 – 5.0 nm with RMS roughness of 1.36 nm. (c) 3D image of Si₃N₄ top surface after CMP with RMS roughness of 0.20 nm and a correlation length of 2.96 nm. (d) 2D image of Si₃N₄ top surface after CMP and scaled to -1.0 – 1.0 nm with RMS roughness of 0.20 nm. Note the different scale bars on (a) and (c).

## 3. Fundamental loss extraction and discussion

The quality factor is a measure of the sharpness of the resonance relative to its central frequency. It represents how well the resonator can store energy and can be written as[11,12]:

$$Q_L = \frac{\omega_0}{\Delta\omega} \qquad (1)$$

The quality factor defined in **Equation 1** is the loaded quality factor. The intrinsic quality factor of the cavity which is directly related to the propagation losses can be written as[13,14]:

$$Q_i = \frac{2Q_L}{1 \pm \sqrt{T_{min}}} \qquad (2)$$

$T_{min}$ is the on-resonance normalized transmission minimum, ± sign is corresponding to undercoupled and overcoupled condition. The schematic of the experimental setup for quality factor measurement and frequency comb generation is shown in **Figure 4**. The resonators we fabricated and measured here have a height of 730 nm, a width of 1500 nm and a bending radius of 115 μm. We measure an intrinsic quality factor of 724,000, corresponding to a propagation loss of 0.42 dB/cm. In **Figure 5(a)**, we show the measured resonance and normalized transmission spectrum over a broad wavelength range. To the best of our knowledge, this is the lowest propagation loss reported to date in a standard PECVD film compatible with foundries.

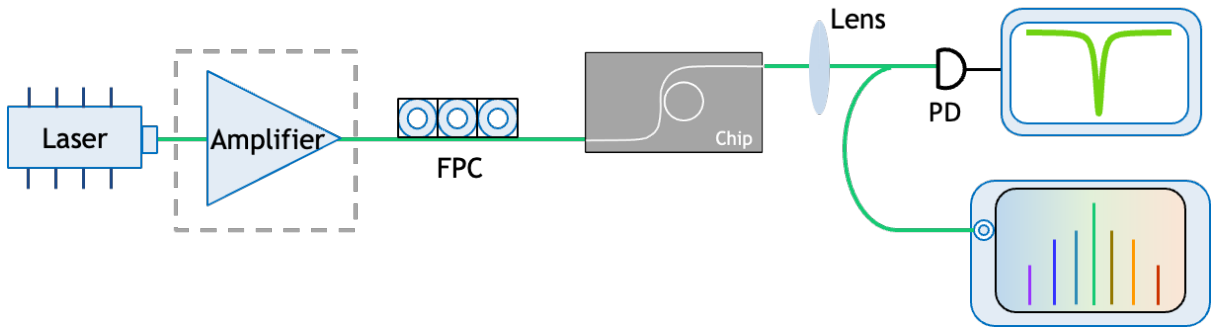

**Figure 4**. Schematic of the experimental setup for measuring transmission spectra and resonator linewidth to characterize the quality factor and generate frequency combs. FPC: fiber polarization controller; PD: photodetector; and OSA: optical spectrum analyzer. Note that amplifier is not needed for transmission measurement.

To minimize both surface scattering losses, as well as bulk loss, we post-process the films with a rapid thermal anneal (RTA). With RTA, we achieve an even higher intrinsic quality factor of more than 1 million, corresponding to a propagation loss of 0.28 dB/cm. RTA has been successfully applied in the microelectronics industry and it has particular relevance for CMOS technology, specifically in steps such as implant annealing, oxidation, and source and drain contact junctions[15,16]. The process reduces loss by driving out the non-bonded atomic and molecular hydrogen trapped in microvoids of the structure and further densifies the films[17,18]. We apply RTA at 800 °C for 5 mins to the cladded devices. In **Figure 5(b)**, we show the measured resonance and normalized transmission spectrum over a broad wavelength range. The thermal budget is below the tolerance of most CMOS electronics and can be used to further reduce losses for devices with microheaters or dopants.

We show that by post-processing foundry-compatible devices with furnace anneal (appropriate for devices with high thermal budget), the propagation loss can be comparable to those fabricated using high temperature, high-stress LPCVD films. Furnace anneal differs from RTA, with higher temperatures (above 1000 °C [19–24]) and longer anneal times (several hours). We anneal cladded devices at 1150 °C in a nitrogen atmosphere for 3 hours and no defects or cracks were observed. We achieve a quality factor of 4.7 million, which corresponds to a propagation loss of 0.06 dB/cm. In **Figure 5(c)**, we show the measured resonance and normalized transmission spectrum over a broad wavelength range.

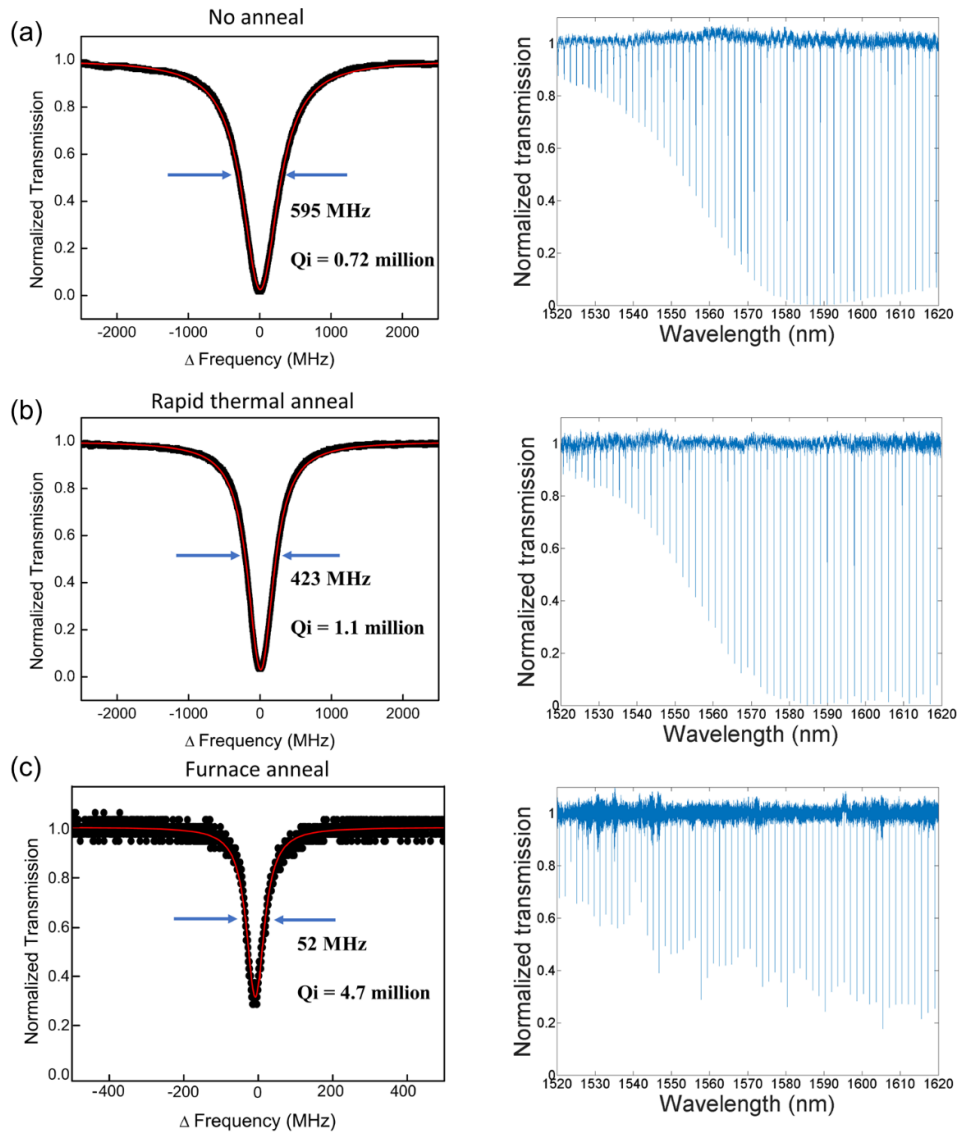

**Figure 5**. (a) Device without annealing shows a measured full width half maximum (FWHM) of 595 MHz around 1600 nm and measured normalized transmission spectrum over a broad wavelength range. (b) Device after rapid thermal anneal shows a measured full width half maximum (FWHM) of 423 MHz around 1600 nm and measured normalized transmission spectrum over a broad wavelength range. (c) Device after furnace anneal shows a measured full width half maximum (FWHM) of 52 MHz around 1600 nm and measured normalized transmission spectrum over a broad wavelength range.

We show that for as-fabricated devices, the bulk losses dominate over the surface scattering loss, and can be as low as 0.33 dB/cm, while for post-fabrication annealed devices, the bulk losses are comparable to the surface scattering loss, and can be as low as 0.04 dB/cm. We extract the loss contributions by comparing the losses between two different structures with different mode overlap with the interfaces. $\eta_1, \eta_2, \eta_3$ are the overlap of the optical field with the waveguide core, the top and bottom surfaces, and sidewalls respectively for the two different waveguide widths[25]. These parameters are calculated using FEM simulations (performed with COMSOL). We also use the Payne-Lacey model[26] to relate scattering loss to the surface's RMS roughness ($\sigma$) and the correlation length ($L_c$), both extracted from the AFM measurements. The method used here to extract the loss contributions is similar to the one used in ref[9]. We find that for complete overlap of the mode with the interfaces, the scattering losses are $\alpha_{top\_scatter} \sim 0.0002$ dB/cm and $\alpha_{bottom\_scatter} \sim 0.0024$ dB/cm at the SiO$_2$/Si$_3$N$_4$ top interface and Si$_3$N$_4$/SiO$_2$ bottom interface, respectively. The estimated surface scattering and bulk loss contributions for different thermal treatments (shown in **Table 1**) are extracted from **Equation 3** and **Equation 4** below:

$$\alpha_{ring1} = \alpha_{bulk\_loss} + \alpha_{top\_scatter} + \alpha_{bottom\_scatter} + \alpha_{sidewalls\_scatter} \tag{3}$$

$$\alpha_{ring2} = \eta_1 \alpha_{bulk\_loss} + \eta_2 (\alpha_{top\_scatter} + \alpha_{bottom\_scatter}) + \eta_3 \alpha_{sidewalls\_scatter} \tag{4}$$

We find that both bulk loss and surface scattering losses are reduced after RTA and furnace anneal, which indicates that the chemical and physical properties of the films are improved by thermal treatment. From Table 1 and Equation 3, if the surface scattering loss were eliminated, one could reduce the propagation loss down to 0.33 dB/cm. By post-processing with RTA at 800 °C, one could reduce the propagation loss to 0.23 dB/cm. The propagation loss can be further reduced if RTA were performed at a higher temperature to break down bonded hydrogen. By post-processing with furnace anneal, one could reduce the propagation loss in these devices to 0.04 dB/cm if the surface scattering loss were eliminated.

**Table 1.** The extracted surface scattering and bulk loss contribution in PECVD film.

|  | Bulk Loss | Surface Scattering Loss | Total Loss |
| --- | --- | --- | --- |
| No Anneal | 0.33 dB/cm | 0.09 dB/cm | 0.42 dB/cm |
| Rapid Thermal Anneal | 0.23 dB/cm | 0.05 dB/cm | 0.28 dB/cm |
| Furnace Anneal | 0.04 dB/cm | 0.02 dB/cm | 0.06 dB/cm |

The structure fabricated without any post-fabrication thermal treatment exhibits a high confinement of 87% and a low propagation loss of 0.42 dB/cm. High confinement is necessary for tailoring the waveguide dispersion to achieve phase matching in nonlinear processes as well as for tighter bends, thus allowing small footprints required in large-scale photonic systems. We compare the confinement factor and propagation loss achieved in this work with other state-of-the-art works realized in foundry compatible PECVD platform without any thermal treatment in **Figure 6**[2,3,5,27–30].

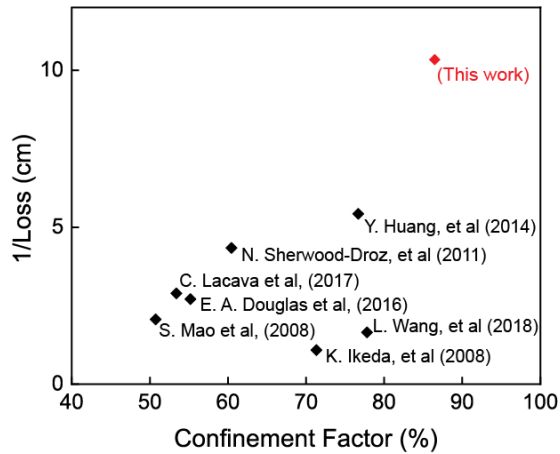

**Figure 6**. Loss and confinement achieved in this work compared with other state-of-the-art works based on PECVD platform. All points including this work are for devices fabricated without any thermal treatment[2,3,5,27–30].

## 4. Dispersion engineering

We show the dispersion of the devices can be tuned by post-processing with furnace anneal. In order to engineer the dispersion, we derive the Sellmeier equations for PECVD $Si_3N_4$ films from ellipsometry performed over 200–1690 nm and 1.7–34 μm wavelength ranges using J.A. Woollam M-2000 and IR-VASE instruments. We show the measured spectra from 200-1750 nm before and after annealing in **Figure 7(a)** and **Figure 7(b)**. We fit the spectra over the wavelength range 300–2000 nm to obtain the following Sellmeier equations for $Si_3N_4$ before and after furnace anneal.

$$n^2_{Si_3N_4}(before\_anneal) = 1 + \frac{2.61\lambda^2}{\lambda^2 - 139.77^2} + \frac{1.11 \times 10^9 \lambda^2}{\lambda^2 - (2.51 \times 10^8)^2}$$

$$n^2_{Si_3N_4}(after\_anneal) = 1 + \frac{2.97\lambda^2}{\lambda^2 - 144.86^2} + \frac{1.57 \times 10^9 \lambda^2}{\lambda^2 - (3.80 \times 10^8)^2}$$

$\lambda$ is in units of nanometer. We show the simulated dispersions based on the Sellmeier equations for silicon nitride resonators with a cross section of 730 nm x 1500 nm and a bending radius of 115 μm before and after annealing in **Figure 7(c)**. The dashed line separates the anomalous group-velocity dispersion (GVD) regime and the normal GVD regime. One can see that the device with the same cross section of 730 nm x 1500 nm exhibits normal GVD before anneal and anomalous GVD after anneal.

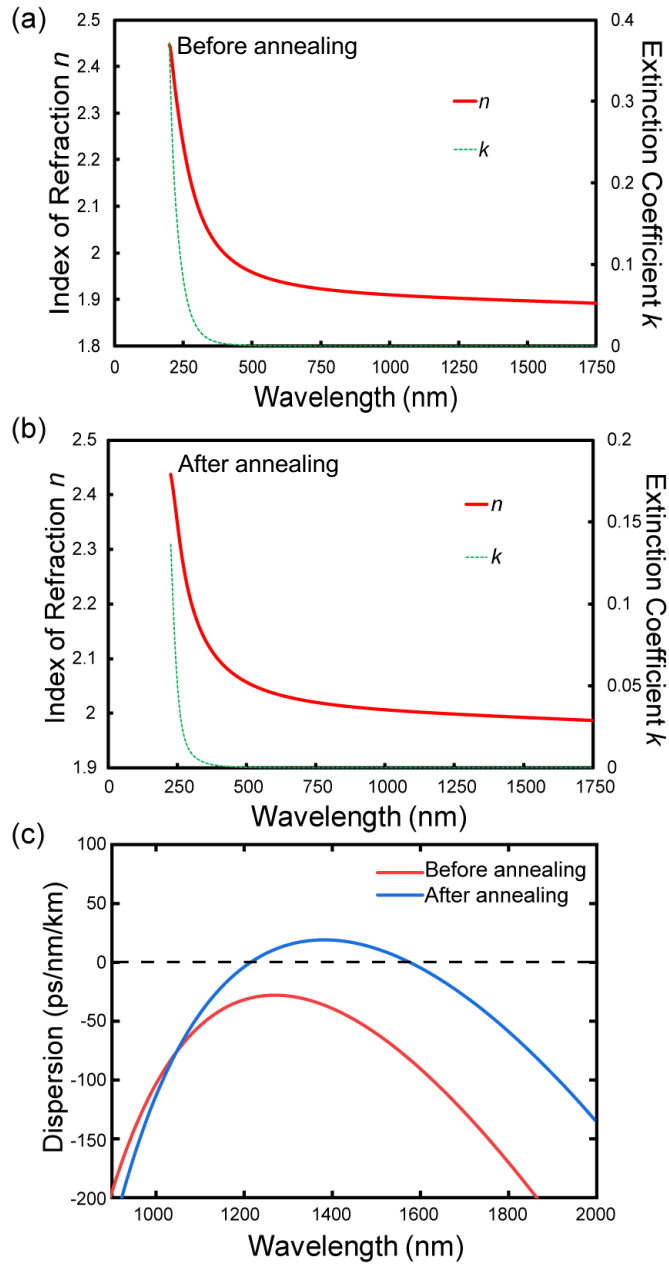

**Figure 7**. (a) Refractive index $n$ and extinction coefficient $k$ for the wavelength range 200–1750 nm before annealing. (b) Refractive index $n$ and extinction coefficient $k$ for the wavelength range 200-1750 nm after annealing. (c) Dispersion simulations for fundamental TE mode of a silicon nitride ring resonator with a cross section of 730 nm × 1500 nm and a bending radius of 115 μm before and after annealing. The dashed line separates the anomalous group-velocity dispersion regime and the normal group-velocity dispersion regime.

## 5. Linear and nonlinear applications

We demonstrate low threshold parametric oscillation and frequency combs generation using foundry compatible devices post-processed with furnace anneal leveraging our ability to engineer the dispersion. We show the evolution of the comb generation process and observe transitions into various comb states in **Figure 8** using a pump wavelength of 1550 nm. As the power in the resonator builds, we see the primary sidebands form at the parametric gain peak due to degenerate four-wave mixing as shown in Figure 8(a). We show the transition into the mini-combs in Figure 8(b) and eventually the broadband frequency combs with an on-chip pump power of 202 mW in Figure 8(c). The parametric oscillation threshold is measured as low as 3 mW, which is close to the theoretical limit of 2.7 mW.

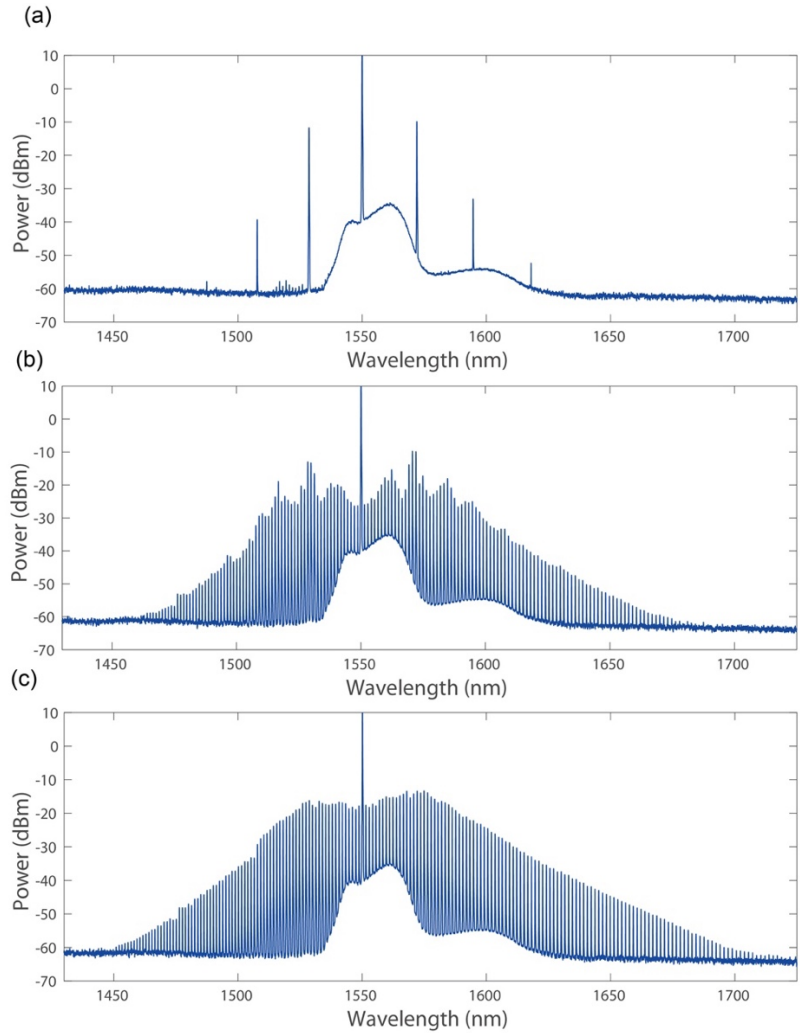

**Figure 8**. Evolution of the frequency comb generation process. (a) Primary sidebands form at the parametric gain peak due to degenerate four-wave mixing. (b) The mini-comb formation. (c) Broadband Kerr frequency comb with an on-chip pump power of 202 mW.

We demonstrate that modal-collapse of a multimode Fabry-Perot laser diode (FPL) can be realized by using the same device. Therefore, we obtain a single-wavelength emission laser thanks to the increased robustness to coupling loss of a FPL[31] and strong feedback of the high quality factor resonator. The system is composed of a commercial single transverse-mode FPL (Thorlabs FPL1001C) and the high quality resonator as shown in **Figure 9**.

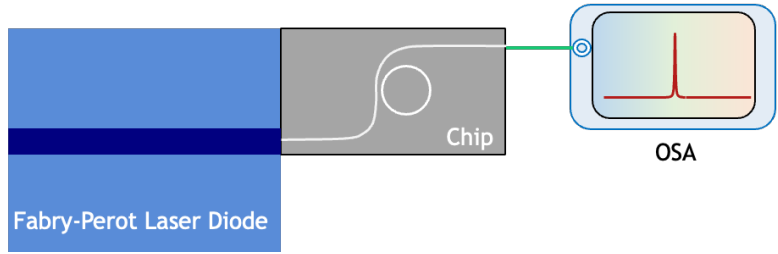

**Figure 9**. Schematic of the experimental setup for lasing measurement. A commercial single transverse-mode Fabry-Perot Laser Diode (Thorlabs FPL1001C) is coupled to the high quality factor resonator. The spectrum of the laser is measured with an optical spectrum analyzer (OSA).

A feedback signal from the high quality factor resonator leads to self-injection locking of the FPL laser resulting in a locked laser with single longitudinal-mode emission and narrow-linewidth. The spectrum of the unlocked free-running laser and the locked laser are shown in **Figure 10**. The side-mode suppression ratio (SMSR) is at least 29 dB and the linewidth is measured below resolution limit of the optical spectrum analyzer. We have calculated the intrinsic linewidth to be in the range of 1 - 10 kHz. For this calculation we have considered the Schawlow–Townes linewidth of the free-running laser and the linewidth reduction due to self-injection locking following a similar procedure as explained in Ref [31]. The coupling structure for our device here is inverse taper and it could be optimized for coupling to FPL, so better SMSRs and even narrower linewidths can be achieved with improved coupling.

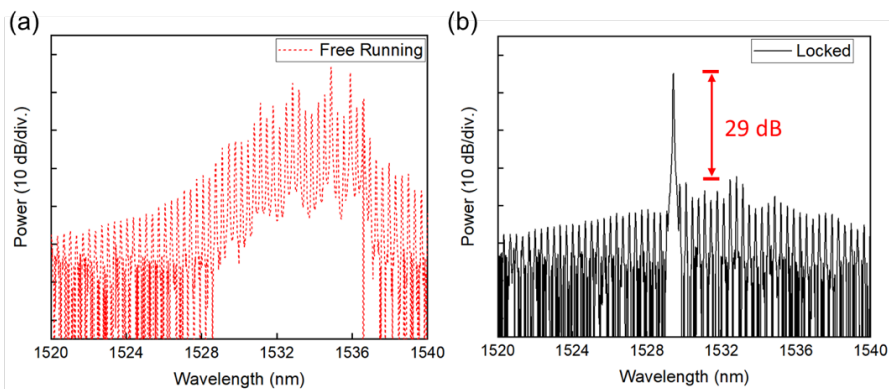

**Figure 10**. (a) Optical spectra of the unlocked free-running laser. (b) Optical spectra of the locked narrow-linewidth laser to the ring resonator. Side-mode suppression ratio (SMSR) is at least 29 dB.

## 6. Conclusion and Discussion

Our work demonstrates the feasibility of obtaining ultra-low loss devices directly from foundries. We show that these foundry compatible devices with or without a simple post-processing step can be used for linear and nonlinear applications where ultra-low loss and dispersion are required. Low threshold parametric oscillation, broadband frequency combs and narrow-linewidth laser are demonstrated. The fundamental limit of loss in our devices is extracted and proved to be comparable with the loss achieved in LPCVD films. Our work provides a promising path for scalable photonic systems based on foundries.

Recently, reactive sputtering silicon nitride films annealed at 400℃ in ambient atmosphere have been shown to achieve propagation losses down to 0.54 dB/cm[32]. Optical frequency combs[32] and hybrid integration with lithium niobate on insulator platforms[33,34] have been successfully demonstrated, which makes the reactive sputtering another promising method for producing low-loss silicon nitride films. Since the losses in reactive sputtering devices are currently limited by scattering from the sidewall roughness rather than H-bond absorption losses[35], these devices could further benefit from the processes and techniques we developed here.

**Acknowledgements**

The authors would like to acknowledge Ron Synowicki from J.A. Woollam Co., the leading manufacturer of spectroscopic ellipsometers for optical properties measurements. Research reported in this work was performed in part at the Cornell NanoScale Science & Technology Facility (CNF), a member of the National Nanotechnology Coordinated Infrastructure (NNCI) supported by National Science Foundation (Grant NNCI-2025233). The authors acknowledge support from the PIPES program funded by DARPA (HR0011-19-2-0014), the PINE program funded by the ARPA-E (DE-AR0000843), and the AFOSR STTR program (FA9550-20-1-0297).